\documentclass[aps,prd,preprintnumbers,notitlepage,superscriptaddress,nofootinbib,longbibliography,10pt]{revtex4-1}

\usepackage{graphicx}
\usepackage[caption=false]{subfig}
\usepackage{amssymb}
\usepackage{amsfonts}
\usepackage{mathtools}
\usepackage{color}
\usepackage[colorlinks=true,linkcolor=magenta,citecolor=cyan]{hyperref}
\usepackage[utf8]{inputenc}
\usepackage[normalem]{ulem}

\newcommand*{\phid}{{\dot \phi}}
\newcommand*{\na}{{\nabla}}

\newcommand*{\Mpld}{{M_{\rm Pl}^2}}
\newcommand*{\chitil}{{\tilde{\chi}}}
\newcommand*{\sgn}{\,{\rm sgn}}

\newcommand*{\ve}[1]{{\rm{\mathbf #1}}}

\begin{document}

\preprint{KOBE-COSMO-20-09}

\title{Theories with limited extrinsic curvature and a nonsingular anisotropic universe}

\author{Yuki Sakakihara}
\email{yukis@mail.sysu.edu.cn}
\affiliation{School of Physics and Astronomy, Sun Yat-sen University, Guangzhou 510275, China}

\author{Daisuke Yoshida}
\email{dyoshida@hawk.kobe-u.ac.jp}
\affiliation{Department of Physics, Kobe University, Kobe 657-8501, Japan}

\author{Kazufumi Takahashi}
\email{ktakahashi@people.kobe-u.ac.jp}
\affiliation{Department of Physics, Kobe University, Kobe 657-8501, Japan}

\author{Jerome Quintin}
\email{jerome.quintin@aei.mpg.de}
\affiliation{Max Planck Institute for Gravitational Physics (Albert Einstein Institute), D-14476 Potsdam, Germany}

\begin{abstract}
We propose a class of theories that can limit scalars constructed from the extrinsic curvature. Applied to cosmology, this framework allows us to control not only the Hubble parameter but also anisotropies without the problem of Ostrogradsky ghost, which is in sharp contrast to the case of limiting spacetime curvature scalars. Our theory can be viewed as a generalization of mimetic and cuscuton theories (thus clarifying their relation), which are known to possess a structure that limits only the Hubble parameter on homogeneous and isotropic backgrounds. As an application of our framework, we construct a model where both anisotropies and the Hubble parameter are kept finite at any stage in the evolution of the universe in the diagonal Bianchi type~I setup. The universe starts from a constant-anisotropy phase and recovers Einstein gravity at low energies. We also show that the cosmological solution is stable against a wide class of perturbation wavenumbers, though instabilities may remain for arbitrary initial conditions.
\end{abstract}

\maketitle

\section{Introduction}

Singularities in the universe~\cite{Penrose:1964wq,Hawking:1966jv,Hawking:1969sw} have been recognized as a problem that demonstrates the failure to describe the Universe in classical Einstein gravity. As is well known, a decelerating and expanding universe inevitably has an initial big bang singularity. Even in an inflationary universe, there is a hard-to-avoid initial singularity~\cite{Borde:2001nh,Yoshida:2018ndv,Numasawa:2019juw}. Since classical Einstein gravity should be a low-energy effective theory of some more fundamental theory of quantum gravity, the presence of singularities is expected to be an artifact of the classical theory, and they should be removed if the effects of quantum gravity are taken into consideration. This has been the hypothesis behind the proposed \emph{limiting curvature} conjecture: there exists a fundamental energy scale, which bounds all physical quantities~\cite{Markov:1982aa,Markov:1987aa,Ginsburg:1988jq}. The idea of the existence of a fundamental energy scale (or length scale \cite{Hossenfelder:2012jw}) is similar to the speed of light in special relativity and the Planck constant in quantum mechanics. This hypothesis has motivated studies of, e.g., a black hole geometry with finite curvature invariants~\cite{Frolov:1989pf,Frolov:1988vj,Morgan:1990yy,Borde:1996df}. Such a non-singular black hole solution was first proposed by Bardeen \cite{bardeen1968non}, which is now known to be a stable solution of Einstein gravity with non-linear electrodynamics~\cite{AyonBeato:2000zs,Moreno:2002gg,Nomura:2020tpc}.

The purpose of the present paper is to propose a new framework to realize the hypothesis of limiting curvature dynamically. Such a theory can be regarded as a candidate of a low-energy effective theory of some unknown theory of quantum gravity.\footnote{In fact, it is known that limiting curvature can be a valid effective-field-theory description of different quantum gravity proposals, such as Loop Quantum Cosmology (LQC), Ho\v{r}ava-Lifshitz gravity, group field theory, etc. As an example, current models of limiting curvature mimetic gravity exactly yield the cosmological background equations of LQC (see, e.g., Refs.~\cite{Bodendorfer:2017bjt,Liu:2017puc,BenAchour:2017ivq,deHaro:2018sqw,deHaro:2018hiq,deCesare:2019pqj,Bezerra:2019bgt,Casalino:2020vhl}, as well as Refs.~\cite{Afshordi:2009tt,Ramazanov:2016xhp,Bodendorfer:2018ptp,deCesare:2018cts} in other quantum gravity contexts).}\ A dynamical realization of the hypothesis, called the limiting curvature theory, was first proposed by Refs.~\cite{Mukhanov:1991zn,Brandenberger:1993ef} in the context of cosmology. The theory was then applied not only to avoid the initial singularity of the Universe~\cite{Moessner:1994jm,Easson:2006jd,Yoshida:2017swb} but also to remove the singularity appearing inside black hole horizons~\cite{Trodden:1993dm,Easson:2002tg, Yoshida:2018kwy}. In the original proposals~\cite{Mukhanov:1991zn,Brandenberger:1993ef}, the authors introduced two scalar fields with a specific potential to limit two spacetime curvature invariants which reduce to the Hubble parameter and its time derivative for a homogeneous and isotropic universe. They found non-singular solutions approaching the de Sitter spacetime at past infinity in the homogeneous and isotropic setup. Regarding the stability, it was shown in Ref.~\cite{Yoshida:2017swb} that the solutions with the curvature invariants of the original proposals are unstable. A stable solution was obtained in the same paper with another choice for the curvature invariants (though bounding the same quantities in the homogeneous and isotropic limit), where the potential and the initial conditions for the scalar fields were fine-tuned. As such, the limiting curvature theory generically exhibits instabilities, which may be associated with the Ostrogradsky ghost~\cite{Woodard:2015zca} due to the presence of higher-order curvature invariants. It should be noted that the assumption that the spacetime is homogeneous and isotropic might be too strong. Indeed, if anisotropies exist, they give a non-negligible contribution to the Friedmann equations. Yet, anisotropies typically tend to diverge in the approach to a spacetime singularity. Moreover, in a collapsing universe, anisotropies are believed to behave chaotically, which is known as the Belinsky-Khalatnikov-Lifshitz (BKL) instability~\cite{Belinsky:1970ew}. Accordingly, the stability against deviations from perfect isotropy is rather non-trivial, and some solutions that are typically stable against inhomogeneities can become unstable when introducing anisotropies~(e.g., \cite{DeFelice:2010hg,Yoshida:2017swb,Pookkillath:2020iqq}).

The original limiting curvature theory \cite{Mukhanov:1991zn,Brandenberger:1993ef} is not the only way to realize the limiting curvature hypothesis. It was pointed out that mimetic gravity~\cite{Chamseddine:2013kea,Chamseddine:2014vna,Sebastiani:2016ras} and cuscuton gravity~\cite{Afshordi:2006ad,Afshordi:2007yx} have a structure that limits the Hubble parameter, and hence they possess non-singular cosmological and black hole solutions~\cite{Chamseddine:2016uef,Chamseddine:2016ktu,Chamseddine:2019pux,Boruah:2018pvq,Quintin:2019orx}. One of the critical differences between mimetic and cuscuton theories (and extensions thereof) is the number of degrees of freedom; mimetic gravity has three degrees of freedom (e.g., \cite{Chaichian:2014qba,Kluson:2017iem,Takahashi:2017pje}), while cuscuton gravity has only two degrees of freedom on a cosmological background (e.g., \cite{Gomes:2017tzd,Lin:2017oow,Chagoya:2018yna,Iyonaga:2018vnu,Mukohyama:2019unx,Gao:2019twq}). In mimetic gravity, the authors of Ref.~\cite{Chamseddine:2016uef} studied an anisotropic universe and found a non-singular Kasner solution.\footnote{Even if singularities are avoided in one anisotropic spacetime, it does not mean singularity resolution is achievable in any arbitrary anisotropic spacetime. For example, mimetic gravity cannot avoid the divergence of anisotropies in an anisotropic Kantowski-Sachs universe \cite{deCesare:2020swb}.}\ However, it is expected that this solution is unstable since we know that cosmological solutions are unstable in a large class of mimetic gravity~\cite{Ramazanov:2016xhp,Ijjas:2016pad,Firouzjahi:2017txv,Zheng:2017qfs,Takahashi:2017pje,Langlois:2018jdg}. On the other hand, cosmological solutions in cuscuton gravity and its extensions can be stable~\cite{Afshordi:2007yx,Boruah:2017tvg,Iyonaga:2018vnu,Iyonaga:2020bmm}, implying that one can construct stable non-singular solutions in cuscuton theories~\cite{Boruah:2018pvq,Quintin:2019orx}.

What we propose in the present paper is that mimetic and cuscuton theories can be understood in a unified framework by reformulating them as limiting curvature theories with respect to the trace of the extrinsic curvature rather than spacetime curvature invariants. In addition, we provide a wider class of theories that can limit desired spatial scalar quantities constructed from the extrinsic curvature. We mention that this model reduces to the framework of spatially covariant theories proposed in Ref.~\cite{Gao:2014soa} (further studied in Refs.~\cite{Gao:2014fra,Fujita:2015ymn,Gao:2018znj,Gao:2019lpz,Gao:2019twq,Gao:2020yzr}) after eliminating the auxiliary scalar fields by using their equations of motion. As an important application of our theory, we explore a non-singular universe with anisotropies. For this purpose, we limit the trace and traceless parts of the extrinsic curvature, which makes both the Hubble parameter and the anisotropies finite.

Our paper is organized as follows. In section~\ref{sec:setup}, we give a general picture of the proposed limiting extrinsic curvature theory. In section~\ref{sec:limK}, we show how to interpret mimetic gravity and cuscuton gravity in the language of limiting extrinsic curvature theory. Also, we demonstrate the similarity between mimetic and cuscuton models by comparing the covariant equations of motion and how the Hubble parameter is kept finite on a homogeneous spacetime. In section~\ref{sec:limKsq}, we study a model where anisotropies are also bounded, which can be regarded as an extension of cuscuton gravity. Based on this model, we construct a Bianchi I spacetime solution with finite Hubble parameter and anisotropies. We also examine the stability of the cosmological solution. Section~\ref{sec:discussion} is devoted to the conclusions and discussion.

\section{General setup for limiting extrinsic curvature}
\label{sec:setup}

We propose a framework of limiting extrinsic curvature theories from an analogy with the limiting spacetime curvature theory proposed in Refs.~\cite{Mukhanov:1991zn,Brandenberger:1993ef}. The original limiting curvature models are based on a Lagrangian density
\begin{equation}
 {\cal L} = \text{(terms independent of $\chi_{k}$)}  + \sum_{k=1}^{n} \chi_{k} I_{k}(R^{\mu}{}_{\nu\rho\sigma}, g_{\mu\nu}, \nabla_{\mu}) - V(\chi_{1}, \chi_{2}, \dots, \chi_{n})\ ,\label{LCT}
\end{equation}
where the $I_{k}$'s are some scalar curvature invariants constructed from the Riemann tensor~$R^{\mu}{}_{\nu\rho\sigma}$, the metric tensor~$g_{\mu\nu}$, and its associated covariant derivative~$\nabla_{\mu}$. Here, the $\chi_{k}$'s are auxiliary scalar fields, whose role is to bound the curvature invariants~$I_{k}$. From the variation of $\chi_{k}$, we obtain a set of equations of motion, which act as constraint equations:
\begin{equation}
 I_{k} = V_{\chi_{k}}(\chi_{1}, \chi_2, \dots, \chi_{n}) \coloneqq \frac{\partial V}{\partial\chi_k}\ .\label{eq:genLimCurvConstr}
\end{equation}
Indeed, by choosing the potential~$V$ so that all of its first derivatives are finite for any configuration of $\chi_{k}$, the curvature invariants~$I_k$ remain finite (see the \hyperref[ssec:limpot]{Appendix} for a detailed discussion). Our proposal is to extend the idea of the limiting curvature theories~\eqref{LCT} by employing the extrinsic curvature tensor~$K_{\mu\nu}$. Thus, we deal with a Lagrangian
\begin{equation}
 {\cal L} = \text{(terms independent of $\chi_{k}$)}+   \sum_{k = 1}^{n} \chi_{k} I_{k}(K_{\mu\nu}, h_{\mu\nu}, D_{\mu}) - V(\chi_1, \chi_{2}, \dots, \chi_{n} )\ ,\label{eq:calLlimitextcurv}
\end{equation}
where now the $I_{k}$'s are some spatial scalars constructed from the extrinsic curvature~$K_{\mu\nu}$, the induced metric~$h_{\mu\nu}$, and the spatial covariant derivative~$D_{\mu}$ with respect to a given space-like foliation~$\Sigma_{t}$. By introducing the unit normal vector to $\Sigma_{t}$ (let us call it $n^\mu$), we can express the induced metric and the extrinsic curvature as
\begin{align}
 h^{\mu\nu} &= g^{\mu\nu} + n^{\mu} n^{\nu}\ , \nonumber\\
 K_{\mu\nu} & = h_{\mu}{}^{\rho} \nabla_{\rho} n_{\nu} = \nabla_{\mu} n_{\nu} + n_{\mu} n^{\rho} \nabla_{\rho} n_{\nu}\ ,
\end{align}
where $n_\mu n^\mu=-1$ and where spacetime indices are raised or lowered with the spacetime metric tensor.

In order to write the action of limiting extrinsic curvature, we need to specify the foliation~$\Sigma_{t}$, i.e., the configuration of the time-like normal vector~$n^\mu$. One way to achieve this is to assume that the theory breaks general covariance, and thus the spacetime foliation~$\Sigma_{t}$ is chosen from the beginning. This is the case of spatially covariant gravity~\cite{Gao:2014soa,Gao:2014fra,Fujita:2015ymn,Gao:2018znj,Gao:2019lpz,Gao:2019twq,Gao:2020yzr}. Another way is to characterize the foliation by a dynamical field. For example, if we regard $n_{\mu}$ as the gradient of some scalar field, $n_{\mu} = -\nabla_{\mu} \phi$, we can say that this is a theory limiting the extrinsic curvature with respect to constant-$\phi$ slices. Since $n_{\mu}$ has to be a unit vector, we need to impose an additional constraint, $\nabla_{\mu}\phi \nabla^{\mu} \phi = -1$, by hand. Similarly, we can regard $n_{\mu}$ itself as a dynamical vector field~$A_{\mu}$, which is normalized according to $A_{\mu} A^{\mu} = -1$.\footnote{To avoid imposing such constraints, one could instead regard $n_{\mu}$ as an automatically-normalized vector such as $n_{\mu} = -\nabla_{\mu}\phi/\sqrt{- \nabla_{\nu} \phi \nabla^{\nu} \phi }$ or $n_{\mu} = A_{\mu}/\sqrt{- A_{\nu} A^{\nu}}$. Each case may define yet another class of limiting extrinsic curvature theories.}\ This means that $A_\mu$ here is nothing but the aether field \cite{Jacobson:2000xp,Eling:2003rd,Eling:2004dk,Jacobson:2004ts,Zlosnik:2006zu,Jacobson:2008aj}. We focus on these two characterizations of a spacetime foliation,
\begin{equation}
 n_{\mu} =
\begin{cases}
 -\nabla_{\mu} \phi &\text{ with } \nabla_{\mu}\phi \nabla^{\mu}\phi = -1\ , \\
 A_{\mu} &\text{ with } A_{\mu}A^{\mu} = -1\ .
\end{cases}
\end{equation}
Adding the Einstein-Hilbert term, the actions of interest can be written explicitly as
\begin{subequations}
\label{generalactions}
\begin{equation}
 S^{\phi} = \int \mathrm{d}^4 x\, \sqrt{-g} \left[ \frac{M_{\text{Pl}}^2}{2} R + \lambda( \nabla_{\mu} \phi \nabla^{\mu} \phi + 1) +   \sum_{k = 1}^{n} \chi_{k} I_{k}(K^{\phi}_{\mu\nu}, h^{\phi}_{\mu\nu}, D^{\phi}_{\mu}) - V(\chi_1, \chi_{2}, \dots, \chi_{n} )\right]
\end{equation}
for $n_\mu=-\na_\mu\phi$ and
\begin{equation}
 S^{A} = \int \mathrm{d}^4 x\, \sqrt{-g} \left[ \frac{M_{\text{Pl}}^2}{2} R + \lambda( A_{\mu} A^{\mu} + 1) +   \sum_{k = 1}^{n} \chi_{k} I_{k}(K^{A}_{\mu\nu}, h^{A}_{\mu\nu}, D^{A}_{\mu}) - V(\chi_1, \chi_{2}, \dots, \chi_{n} )\right]
\end{equation}
\end{subequations}
for $n_\mu=A_\mu$. Here, the superscripts~$\phi$ and $A$ on $K_{\mu\nu}$, $h_{\mu\nu}$, and $D_{\mu}$ mean that these quantities are defined with respect to constant-$\phi$ hypersurfaces and those normal to $A_{\mu}$, respectively. The role of the term proportional to $\lambda$ in the Lagrangian densities is to enforce the constraints $\nabla_\mu\phi\nabla^\mu\phi=-1$ and $A_\mu A^\mu=-1$. As such, $\lambda$ is a Lagrange multiplier. The above actions thus represent the general framework of limiting extrinsic curvature theories that we propose in this paper. As we will see in the next section, $S^{\phi}$ and $S^{A}$ can be regarded as extensions of mimetic and cuscuton gravity, respectively. Hence, in what follows, we refer to models constructed in the form of $S^\phi$ as `mimetic-type' theories and to those constructed in the form of $S^A$ as `cuscuton-type' theories.

Let us apply this framework to non-singular cosmology. We first focus on a flat Friedmann-Lema{\^i}tre-Robertson-Walker~(FLRW) spacetime,
\begin{equation}
 g_{\mu\nu}\mathrm{d}x^{\mu}\mathrm{d}x^{\nu} = - N(t)^2 \mathrm{d}t^2 + a(t)^2 \left( \mathrm{d}x^2 + \mathrm{d}y^2 + \mathrm{d}z^2\right)\ ,
\end{equation}
where $a(t)$ is the scale factor and $N(t)$ is the lapse function. The Hubble parameter~$H$ is defined by $H \coloneqq \dot{a}/(Na)$, with a dot denoting the time derivative. The question is then what should be chosen for the scalar functions~$I_k$ to avoid divergence in the Hubble parameter~$H$. The simplest example would be the case where we limit the trace of the extrinsic curvature, $K=K^\mu{}_\mu$, which corresponds to the Hubble parameter as
\begin{equation}
 \left.K\right|_{\text{FLRW}} = 3H \ .
\end{equation}
Since the trace of the extrinsic curvature in general satisfies $K=K^\mu{}_\mu \approx \nabla^\mu n_\mu$, where the symbol~$\approx$ represents equality under the condition~$n_{\mu}n^{\mu} = -1$, let us fix $n=1$ (i.e., we bound a single extrinsic curvature invariant) and take $I_{1} = \nabla^{\mu} n_{\mu}$ for our purpose. Depending on the choice of the normal vector, we find two types of theories:
\begin{subequations}
\label{Hactions}
\begin{equation}
 S^{\phi,H}=\int \mathrm{d}^4 x\, \sqrt{-g}\left[\frac{\Mpld}{2}R+\lambda(\nabla_\mu\phi \nabla^\mu\phi+1)-\chi\Box\phi-V(\chi) \right]
 \label{SphiH}
\end{equation}
for $n_\mu=-\nabla_\mu \phi$, where $\Box \coloneqq g^{\mu\nu}\nabla_\mu\nabla_\nu$ is the d'Alembertian; and 
\begin{equation}
 S^{A,H}=\int \mathrm{d}^4 x\, \sqrt{-g}\left[\frac{\Mpld}{2}R+\lambda(A_\mu A^\mu +1)+\chi \nabla^\mu A_\mu-V(\chi) \right]
 \label{SAH}
\end{equation}
\end{subequations}
for $n_\mu=A_\mu$. In the next section, we will see that the former action is equivalent to mimetic gravity with an extension, while the latter is equivalent to cuscuton gravity.

Anisotropies are also problematic when we seek a model avoiding the initial singularity of the Universe. This is because anisotropies dominate the universe at early times since their energy density scales as $a^{-6}$. To avoid such a divergence of anisotropies, in addition to limiting the Hubble parameter, we also limit anisotropies by making use of the mechanism of limiting extrinsic curvature. Here, we consider the diagonal Bianchi I spacetime,
\begin{equation}
 g_{\mu\nu}\mathrm{d}x^{\mu}\mathrm{d}x^{\nu} = - N(t)^2 \mathrm{d}t^2 + a(t)^2 \left( e^{2\beta_{+}(t) + 2 \sqrt{3} \beta_{-}(t)}\mathrm{d}x^2 + e^{2 \beta_{+}(t) -2 \sqrt{3} \beta_{-}(t)}\mathrm{d}y^2 + e^{-4 \beta_{+}(t)}\mathrm{d}z^2\right)\ .\label{223903_24Feb20}
\end{equation}
The following combination of the extrinsic curvature characterizes the overall anisotropy,
\begin{equation}
 K^\mu{}_\nu K^\nu{}_\mu-\frac{1}{3}K^2=6 \Sigma^2 \ ,
\end{equation}
where $\Sigma^2 \coloneqq  \sigma_{+}^2 + \sigma_{-}^2$ and $\sigma_{\pm}\coloneqq  \dot{\beta}_{\pm}/N$. We shall call $\Sigma\coloneqq \sqrt{\Sigma^2}$ the anisotropy parameter. Thus, a model with limiting Hubble parameter and limiting anisotropy parameter can be obtained by fixing $n=2$ (i.e., we bound two extrinsic curvature invariants) and choosing $I_1$ and $I_2$ as follows,\footnote{In order to make $H$ and $\Sigma$ finite, we could instead consider a theory limiting only $K^\mu{}_\nu K^\nu{}_\mu$ since $K^\mu{}_\nu K^\nu{}_\mu=3H^2+6 \Sigma^2$ and since $H^2$ and $\Sigma^2$ are both positive semi-definite. One might think that this model is simpler and sufficient to avoid the divergence of both the Hubble parameter and anisotropies at the same time. However, as far as we investigated, it is impossible to find a theory of this type that recovers Einstein gravity at low energies and that has a homogeneous non-singular spacetime solution in the asymptotic past.}
\begin{align}
 I_{1} &= K^2 \approx (\nabla^{\mu} n_{\mu})^2\ , \nonumber\\
 I_{2} &= K^{\mu}{}_{\nu}K^{\nu}{}_{\mu} - \frac{1}{3}K^2 \approx \nabla^\mu n_\nu \nabla^\nu n_\mu-\frac{1}{3}(\nabla^\mu n_\mu)^2\ ,
\end{align}
where the symbol $\approx$ represents again the equality under the condition $n_{\mu}n^{\mu} = -1$. Hence, we have two versions of this limiting anisotropy construction as
\begin{subequations}
\label{Ssigma}
\begin{equation}
 S^{\phi,\Sigma}=\int \mathrm{d}^4 x\, \sqrt{-g}\left[\frac{\Mpld}{2}R+\lambda(\nabla_\mu\phi \nabla^\mu\phi+1)-\chi_1 \Box\phi+\chi_2 \left(\nabla^\mu\nabla_\nu\phi\nabla^\nu\nabla_\mu\phi-\frac{1}{3}(\Box \phi)^2\right)-V(\chi_1,\chi_2) \right]
 \label{223155_24Feb20}
\end{equation}
for $n_\mu=-\nabla_\mu \phi$ and  
\begin{equation}
 S^{A,\Sigma}=\int \mathrm{d}^4 x\, \sqrt{-g}\left[\frac{\Mpld}{2}R+\lambda(A_\mu A^\mu +1)+\chi_1 (\nabla^\mu A_\mu)^2 +\chi_2\left(\nabla^\mu A_\nu\nabla^\nu A_\mu-\frac{1}{3}(\nabla^\mu A_\mu)^2\right)-V(\chi_1,\chi_2) \right]
 \label{223405_24Feb20}
\end{equation}
\end{subequations}
for $n_\mu=A_\mu$, as before. 
To see the explicit relation between our theories and mimetic/cuscuton models, we discuss both mimetic- and cuscuton-type theories in section \ref{sec:limK}. However, since we know that a large class of mimetic gravity, including 
the model~\eqref{223155_24Feb20}, is plagued by ghost/gradient instabilities~\cite{Ramazanov:2016xhp,Ijjas:2016pad,Firouzjahi:2017txv,Zheng:2017qfs,Takahashi:2017pje,Langlois:2018jdg},\footnote{Inclusion of higher-order curvature invariants, moreover, leads to the appearance of Ostrogradsky ghosts (see also Ref.~\cite{Takahashi:2017pje}).} we focus on the cuscuton-type theory~\eqref{223405_24Feb20} in section \ref{sec:limKsq}.

\section{Limiting $K$ models}
\label{sec:limK}

In this section, we investigate properties of the theories described by \eqref{Hactions}, which can limit the trace of the extrinsic curvature~$K$. In section~\ref{sec:limKA}, we demonstrate that the actions~\eqref{SphiH} and \eqref{SAH} are respectively equivalent to those of mimetic and cuscuton gravity. Then, we derive the covariant equations of motion from the actions \eqref{Hactions} in section \ref{sec:limKB}. Finally, we investigate cosmological solutions in section \ref{ssec:homogeneous}.

\subsection{Relation to mimetic and cuscuton models}
\label{sec:limKA}

The limiting $K$ models are useful to avoid the initial divergence of the Hubble parameter. Here, we show the equivalence of the limiting $K$ models we proposed in the previous section with mimetic and cuscuton gravity.

\subsubsection{Mimetic gravity}

First, let us focus on the action~\eqref{SphiH}, which we reprint below for convenience:
\[
 S^{\phi, H} = \int \mathrm{d}^4 x \,\sqrt{-g} \left[ \frac{\Mpld}{2} R + \lambda ( \nabla_{\mu}\phi \nabla^{\mu}\phi + 1) - \chi \Box \phi - V(\chi)\right] \ .
\]
Let us assume $\partial^2V/\partial\chi^2\ne 0$ so that the equation of motion,
\begin{equation}
 \frac{1}{\sqrt{-g}} \frac{\delta S^{\phi,H}}{\delta \chi} = -\Box \phi - \frac{\partial V}{\partial\chi}= 0\ ,
\end{equation}
can be solved for $\chi$, namely, $\chi = \chi (\Box \phi)$. Then, eliminating the auxiliary field~$\chi$ from the action, we obtain
\begin{equation}
 S^{\phi, H} = \int \mathrm{d}^4 x \,\sqrt{-g} \left[ \frac{\Mpld}{2} R + \lambda ( \nabla_{\mu}\phi \nabla^{\mu}\phi + 1) + f(\Box \phi) \right] \ .
\end{equation}
Here, $f$ is the Legendre transformation of $V$ defined by
\begin{equation}
 f(\Box \phi) \coloneqq -\chi(\Box \phi) \Box \phi - V(\chi(\Box \phi)) \ .
\end{equation}
This is nothing but mimetic gravity with an $f(\Box \phi)$ extension~\cite{Chamseddine:2016uef,Chamseddine:2016ktu} (a model of this form was first studied in Ref.~\cite{Chamseddine:2014vna}).

\subsubsection{Cuscuton gravity}

Next, let us focus on the action \eqref{SAH},
\[
 S^{A,H}=\int \mathrm{d}^4 x \,\sqrt{-g} \left[\frac{\Mpld}{2}R+\lambda(A_\mu A^\mu +1)+\chi \nabla^\mu A_\mu-V(\chi) \right] \ .
\]
We take the variation of the action with respect to $A_\mu$ and obtain
\begin{equation}
 A_\mu=\frac{1}{2\lambda}\nabla_\mu\chi \ .\label{eq:Amunablamuchi}
\end{equation}
By taking into account the normalization of the vector field, which is obtained by variation with respect to $\lambda$, 
\begin{equation}
 A_{\mu}A^{\mu}+1=0 \ ,\label{eq:Amuconstr}
\end{equation}
we find\footnote{By construction, $A_\mu$ is ensured to be time-like, as seen from \eqref{eq:Amuconstr}. Thus from \eqref{eq:Amunablamuchi}, $\nabla_\mu\chi$ is also necessarily time-like. Consequently, $\sqrt{-\nabla_\mu\chi\nabla^\mu\chi}$ always yields a real number in this setup.}
\begin{equation}
 \lambda=\pm\frac{1}{2}\sqrt{-\nabla_\mu \chi\nabla^\mu\chi} \ ,
\end{equation}
and therefore,
\begin{equation}
 A_\mu = \frac{\pm \nabla_\mu \chi}{\sqrt{-\nabla_\nu\chi\nabla^\nu \chi}} \ ,
\end{equation}
where the $\pm$ sign is in the same order. If we substitute this expression into the action, we get
\begin{equation}
 S^{A,H}=\int \mathrm{d}^4 x \,\sqrt{-g}\left[\frac{\Mpld}{2}R\pm\sqrt{-\nabla_\mu\chi\nabla^\mu\chi}-V(\chi) \right]   \ .
\end{equation}
This is nothing but cuscuton gravity\footnote{The Lagrangian of cuscuton gravity usually has a kinetic term of the form~$\pm \mu^2 \sqrt{-\nabla^\mu \chi \nabla_\mu \chi}$, but the coefficient~$\mu^2$ can be absorbed without loss of generality if we define $\chitil\coloneqq \mu^2 \chi$.} with an arbitrary potential \cite{Afshordi:2006ad,Afshordi:2007yx}. Let us mention that the relation between the above action for cuscuton gravity and the limiting extrinsic curvature action \eqref{SAH} was already observed in Refs.~\cite{Afshordi:2009tt,Afshordi:2010eq,Bhattacharyya:2016mah} (in addition to the link with Ho\v{r}ava-Lifshitz gravity and the Einstein-aether theory). However, the link to a wider class of limiting extrinsic curvature theories and the correspondence with mimetic gravity were not realized at the time.

\subsection{Covariant equations of motion}
\label{sec:limKB}

We saw that both mimetic gravity and cuscuton gravity can be understood as a kind of limiting extrinsic curvature theory. They actually have a similar structure at the level of their equations of motion. To see the similarities and the differences, here we show the covariant equations of motion for the mimetic-type and the cuscuton-type limiting $K$ models.

The equations of motion for the mimetic-type model~\eqref{SphiH} with respect to $\lambda$, $\chi$, and $\phi$ are
\begin{subequations}
\begin{align}
 \nabla_\mu\phi \nabla^\mu\phi +1&= 0 \ , \label{limKmimEOM1} \\
 -\Box\phi-\frac{\partial V}{\partial\chi} &= 0 \ , \label{limKmimEOM2} \\
 -\nabla_\mu(2\lambda \nabla^\mu\phi+\nabla^\mu\chi)&=0 \ , \label{lambda_mim}
\end{align}
\end{subequations}
respectively. The second equation implies that the trace of the extrinsic curvature, $K=-\Box \phi$, can be bounded if we choose a potential whose derivative, $\partial V/\partial\chi$, does not diverge at any value of $\chi$. We can integrate the last equation~\eqref{lambda_mim} by introducing a divergenceless vector~$u_\mu$ (i.e., $\nabla^\mu u_\mu=0$, so $u_\mu$ is akin to an integration constant) as
\begin{equation}
 \lambda=\frac{1}{2}(\nabla^\mu\chi+u^\mu)\nabla_\mu \phi\ . \label{int_lambda_mim}
\end{equation}
Taking into account these equations, we find the equation of motion for gravity,
\begin{equation}
 G_{\mu\nu}\coloneqq  R_{\mu\nu}-\frac{1}{2}g_{\mu\nu}R=\frac{1}{\Mpld}\left\{T_{\mu\nu}+g_{\mu\nu}\left[\nabla^\rho \chi \nabla_\rho\phi-V(\chi)\right]-2\nabla_{(\mu}\chi\nabla_{\nu)}\phi- \nabla_\mu\phi\nabla_\nu\phi (\nabla^\rho\chi+u^\rho)\nabla_\rho\phi\right\} \ ,
 \label{limKmimEinstein}
\end{equation}
where round brackets in the spacetime indices denote symmetrization, i.e., $\nabla_{(\mu}\chi\nabla_{\nu)}\phi\coloneqq (\nabla_{\mu}\chi\nabla_{\nu}\phi+\nabla_{\nu}\chi\nabla_{\mu}\phi)/2$. Here,
\begin{equation}
 T_{\mu\nu}=-\frac{2}{\sqrt{-g}}\frac{\delta S_{\rm matter}}{\delta g^{\mu\nu}}
\end{equation}
is the energy-momentum tensor of any additional matter, which is assumed to be minimally coupled to gravity.

On the other hand, the equations of motion for the cuscuton-type model~\eqref{SAH} with respect to $\lambda$, $\chi$, and $A_\mu$ are
\begin{subequations}
\begin{align}
 A_\mu A^\mu+ 1&=0 \ ,\label{193317_8Mar20} \\
 \nabla^\mu A_\mu-\frac{\partial V}{\partial\chi}&=0 \ ,\label{193328_8Mar20} \\
 2\lambda A^\mu-\nabla^\mu\chi &= 0 \ , \label{lambda_cus}
\end{align}
\end{subequations}
respectively. Since the trace of the extrinsic curvature is written as $K=\nabla^\mu A_\mu$, it can again be bounded by an appropriate choice of the potential. In this case, we do not need to integrate the last equation~\eqref{lambda_cus} and simply obtain
\begin{equation}
 \lambda=-\frac{1}{2}\nabla^\mu \chi A_\mu \ .
\end{equation}
This can be applied to the equation of motion for gravity, and we have 
\begin{equation}
  G_{\mu\nu}=\frac{1}{\Mpld}\left\{T_{\mu\nu}+g_{\mu\nu}\left[-\nabla^\rho \chi A_\rho-V(\chi)\right]+2\nabla_{(\mu}\chi A_{\nu)}+A_\mu A_\nu \nabla^\rho \chi A_\rho \right\}\ .\label{193524_8Mar20}
\end{equation}

It is straightforward to see that the equations of motion for the cuscuton-type model coincide with those for the mimetic-type model if we replace $A_\mu$ with $-\nabla_\mu \phi$, except for the appearance of $u_\mu$ in \eqref{limKmimEinstein}. As shown in \eqref{int_lambda_mim}, this $u_\mu$ is nothing but an integration constant originating from the covariant derivative in \eqref{lambda_mim} for the mimetic case. This integration constant is the origin of the matter-like contribution pointed out in Ref.~\cite{Chamseddine:2013kea}. Note that, from the corresponding equation~\eqref{lambda_cus} for the cuscuton case, one can determine $\lambda$ without the ambiguity of an integration constant. Given this similarity between the mimetic and cuscuton models, it is expected that any solution in the mimetic theory reduces to a solution in the cuscuton theory in the limit~$u_\mu\to 0$. Nevertheless, the stability of the solutions can be different due to the different number of degrees of freedom in the two theories. It should also be noted that the additional mode solution that $u_\mu$ introduces in the mimetic-type model makes it manifest that the theory generally has one more degree of freedom compared to the cuscuton-type model.

\subsection{Homogeneous spacetime}
\label{ssec:homogeneous}

We pointed out that solutions found in mimetic gravity should also appear in the context of cuscuton gravity. Along this line, we see that this is indeed true for the diagonal Bianchi I universe described by the metric~\eqref{223903_24Feb20} in the absence of the mimetic matter field. In this subsection, we consider the vacuum case with no additional matter fields.

\subsubsection{Mimetic gravity}

To be consistent with a homogeneous spacetime, we assume
\begin{equation}
  \chi = \chi(t)\ , \qquad  \phi = \phi(t)\ , \qquad \lambda = \lambda(t) \ . \label{limKmim_scalars}
\end{equation}
We use the covariant equations of motion~\eqref{limKmimEOM1}, \eqref{limKmimEOM2}, and \eqref{limKmimEinstein} to study the spacetime dynamics.\footnote{Instead, one can substitute \eqref{223903_24Feb20} and \eqref{limKmim_scalars} into the action~\eqref{SphiH} and then vary it with respect to relevant variables to derive field equations. In doing so, one should choose $N=1$ after the variation, as the equation of motion for $N$ cannot be reproduced from the other components of the field equations~\cite{Motohashi:2016prk}.}\ In what follows, we choose the $N=1$ gauge. From \eqref{limKmimEOM1} and \eqref{limKmimEOM2}, we obtain
\begin{subequations}
\begin{align}
 \phid&=\pm 1\ , \label{limKmimEOM1B} \\
 H&=\frac{1}{3}\frac{\partial V}{\partial\chi}\ . \label{limKmimEOM2B}
\end{align}
\end{subequations}
It should be noted that \eqref{limKmimEOM2B} is essential for limiting the Hubble parameter. By choosing $V$ in such a way that $\partial V/\partial\chi$ is finite, the range of the Hubble parameter is also restricted to a finite energy interval. In what follows, we choose the plus branch of \eqref{limKmimEOM1B} so that $\phi$ behaves as our clock. The evolution of the Hubble parameter and the anisotropies is determined from the modified Einstein equation~\eqref{limKmimEinstein}. The anisotropies $\sigma_{\pm} = \dot{\beta}_{\pm}$ can be obtained as
\begin{equation}
 \sigma_\pm=\frac{\sigma_\pm^{(0)}}{a^{3}} \ ,\label{192834_8Mar20}
\end{equation}
where $\sigma_\pm^{(0)}$ are constants determined by the initial conditions. Then, the Friedmann equation is written as
\begin{equation}
 H^2-\frac{\Sigma^2_0}{a^{6}}=\frac{1}{3\Mpld}\left[V(\chi)-u^0\right]\ , \label{limKmimEinsteinB}
\end{equation}
with $\Sigma^2_0\coloneqq \left(\sigma_+^{(0)}\right)^2+\left(\sigma_-^{(0)}\right)^2$. Here, $\chi$ in the right-hand side should be understood as a function of $H$ through \eqref{limKmimEOM2B}. Provided that $u^\mu=u^\mu(t)$, we have $\nabla_\mu u^\mu=\dot{u}^0+3Hu^0=0$, namely, $u^0\propto a^{-3}$. This implies that the second term in the right-hand side of \eqref{limKmimEinsteinB} plays the role of pressureless dust as pointed out in Ref.~\cite{Chamseddine:2013kea}.

\subsubsection{Cuscuton gravity}

In the case of cuscuton gravity, we assume
\begin{equation}
  \chi = \chi(t), \qquad  A_\mu = (A_0(t),{\bf 0}), \qquad \lambda = \lambda(t)\ .
\end{equation}
We choose again the $N=1$ gauge after variation. The equations of motion~\eqref{193317_8Mar20} and \eqref{193328_8Mar20} read
\begin{subequations}
\begin{align}
 A_0&=\pm 1\ , \\
 H&=\frac{1}{3}\frac{\partial V}{\partial\chi}  \ ,
\end{align}
\end{subequations}
and we choose the minus branch for $A_0$ to make $A^\mu$ future-directed. The Friedmann equation is given by
\begin{equation}
 H^2-\frac{\Sigma^2_0}{a^6}=\frac{1}{3\Mpld}V(\chi)
\end{equation}
from \eqref{193524_8Mar20}. Note that the equations of motion for the anisotropies are the same as those in mimetic gravity, which results in the solutions~\eqref{192834_8Mar20}.

\vskip 3mm
We find that both models have the limiting curvature feature with respect to the Hubble parameter if we choose an appropriate function for the potential. The only difference between these two gravity theories is the existence of a matter-like contribution coming from the integration constant of one of the equations of motion in the mimetic theory, but no such contribution exists in cuscuton gravity. It is obvious that such a difference is coming from the definition of the normal vector; the normal vector~$n^\mu$ is chosen to be the {\it derivative} of a scalar field in the mimetic case, which is not the case in cuscuton gravity. These models are successful in limiting the Hubble parameter, but it should be noted that the anisotropies blow up as $ a \rightarrow 0$ as one can see from \eqref{192834_8Mar20}.

\section{Limiting anisotropy models}
\label{sec:limKsq}

We showed that the two models that limit the trace of the extrinsic curvature have the property of limiting the Hubble parameter. In addition, we explicitly saw that, at the background level, the homogeneous anisotropic universe in mimetic gravity behaves completely in the same way as in cuscuton gravity when the mimetic dust contribution, coming from an integration constant, is turned off. Now, we attempt to cure the divergent behavior of the anisotropies in the early universe by introducing a potential limiting anisotropies. In what follows, we restrict ourselves to the cuscuton-type limiting anisotropy model~\eqref{223405_24Feb20} since the mimetic-type model~\eqref{223155_24Feb20} is in general plagued by instabilities~\cite{Ramazanov:2016xhp,Ijjas:2016pad,Firouzjahi:2017txv,Zheng:2017qfs,Takahashi:2017pje,Langlois:2018jdg}. Indeed, this instability originates from a scalar degree of freedom, which is expected to be absent in the cuscuton-type theory on a cosmological background (e.g., \cite{Gomes:2017tzd,Iyonaga:2018vnu,Mukohyama:2019unx,Gao:2019twq}). As an example, a non-singular homogeneous and isotropic bouncing background in mimetic gravity was shown to have instabilities~\cite{Ijjas:2016pad}, while non-singular bouncing backgrounds in cuscuton gravity have shown no instability \cite{Boruah:2018pvq,Quintin:2019orx}, at the level of linear inhomogeneous perturbations.

\subsection{Covariant equations of motion for cuscuton-type limiting anisotropy model}

We recall the cuscuton-type limiting anisotropy model~\eqref{223405_24Feb20}, which we discuss from now on,
\[
 S^{A,\Sigma}=\Mpld\int \mathrm{d}^4 x \, \sqrt{-g} \left[\frac{R}{2}+\lambda(A_\mu A^\mu +1)+ \chi_1 (\nabla^\mu A_\mu)^2 +\chi_2 \left(\nabla^\mu A_\nu \nabla^\nu A_\mu-\frac{1}{3}(\nabla^\mu A_\mu)^2 \right)-\mu^2V(\chi_1, \chi_2) \right] \ ,
\]
where we rescaled the fields and the potential as $\lambda \rightarrow M_{\text{Pl}}^2 \lambda, \chi_{k} \rightarrow M_{\text{Pl}}^2 \chi_{k}$ ($k=1,2$), and $V \rightarrow M_{\text{Pl}}^2 \mu^2 V$. The equations of motion derived from this action are
\begin{subequations}
\begin{align}
 &A_\mu A^\mu+1 =0\ ,\\
 &\lambda A_\mu=\nabla_\mu \left[\left(\chi_1-\frac{1}{3}\chi_2 \right)\nabla^\nu A_\nu\right]+\nabla^\nu(\chi_2 \nabla_\mu A_\nu )\ , \label{095549_5Mar20} \\
 &(\nabla^\mu A_\mu)^2=\mu^2  V_{\chi_1}\ ,\label{095550_5Mar20} \\
 &\nabla^\mu A_\nu\nabla^\nu A_\mu-\frac{1}{3}(\nabla^\mu A_\mu)^2
 =\mu^2  V_{\chi_2}\ ,\label{095557_5Mar20} 
\end{align}
\end{subequations}
where we recall the shorthand notation $V_{\chi_k} \coloneqq \partial V/\partial \chi_k$ ($k=1,2$), and
\begin{equation}
 \Mpld G_{\mu\nu}=T_{\mu\nu}+T^{\rm (lim)}_{\mu\nu}\ ,
 \label{lim-aniso_EOM-gmn}
\end{equation}
where $T_{\mu\nu}$ is the energy-momentum tensor of additional, minimally-coupled matter fields, and $T_{\mu\nu}^{\rm (lim)}$ is the effective energy-momentum tensor coming from the limiting-curvature part,
\begin{align}
 \frac{T^\mathrm{(lim)}_{\mu\nu}}{\Mpld}=&~g_{\mu\nu}\left\{\left(\chi_1-\frac{1}{3}\chi_2\right)(\nabla^\alpha A_\alpha)^2+\chi_2(\nabla^\alpha A_\beta\nabla^\beta A_\alpha)-\mu^2V-2\nabla^\alpha\left[\left(\chi_1-\frac{1}{3}\chi_2\right)A_\alpha\nabla^\beta A_\beta\right]\right\}\nonumber\\
  & +2A_{(\mu}\nabla_{\nu)}\left[\left(\chi_1-\frac{1}{3}\chi_2\right)\nabla^\alpha A_\alpha\right]
   + 2 \nabla^\alpha \left[\chi_2 \nabla_\alpha A_{(\mu}\right]A_{\nu)}-2\nabla^\alpha \left[\chi_2 A_\alpha\nabla_{(\mu} A_{\nu)}\right]\nonumber\\
  &+2\chi_2\nabla_\alpha A_{(\mu}\left[\nabla^\alpha A_{\nu)}-\nabla_{\nu)}A^\alpha\right]\ .
\end{align}
Here, we used \eqref{095549_5Mar20} to eliminate $\lambda$.

\subsection{Evolution of Bianchi I spacetime}

By substituting the Bianchi I spacetime ansatz~\eqref{223903_24Feb20} into the action~\eqref{223405_24Feb20}, we obtain a Lagrangian written in terms of $N,\,a,\,\beta_\pm,\,\lambda,\,A_\mu,\,\chi_1,\,\chi_2$, from which the equations of motion are derived. We choose the $N=1$ gauge and find $A_0=\pm 1$ from the variation with respect to $\lambda$. We take the minus branch from here on as in the previous section. Then, the gravitational equations of motion are
\begin{subequations}
\begin{align}
 (1-3\chi_1)H^2-(1+2\chi_2) \Sigma^2&=\frac{\rho}{3\Mpld}+\frac{\mu^2 V}{3} \label{friedmann_K2}\ , \\
 \frac{1}{a^3}\frac{\mathrm{d}}{\mathrm{d}t}\left[a^3 (1+2\chi_2)\sigma_{\pm}\right]&=\frac{p_{\pm}}{3\Mpld}\ , \\
 \frac{\mathrm{d}}{\mathrm{d}t}\left[(1-3\chi_1)H\right]+3(1+2\chi_2) \Sigma^2&=-\frac{1}{2\Mpld}(\rho+p)\ ,
\end{align}
\end{subequations}
where we decomposed the matter energy-momentum tensor as
\begin{equation}
 T^\mu{}_{\nu}=
\begin{pmatrix}
 -\rho &&&\\
 & p+\frac{p_+}{3}+\frac{p_-}{\sqrt{3}}& &\\ 
 & & p+\frac{p_+}{3}-\frac{p_-}{\sqrt{3}}& \\
 & & &  p-\frac{2p_+}{3}
\end{pmatrix}\ .
\end{equation}
Note that we used the equation determining $\lambda$,
\begin{equation}
 \lambda= 3\chi_2 (H^2+2 \Sigma^2)+\frac{\mathrm{d}}{\mathrm{d}t}\left[(\chi_2-3\chi_1)H\right] \ .\label{183200_24Feb20}
\end{equation}
Also, equations~\eqref{095550_5Mar20} and \eqref{095557_5Mar20} become
\begin{equation}
 H^2=\frac{\mu^2 V_{\chi_1}}{9} \ ,\qquad 
 \Sigma^2=\frac{\mu^2 V_{\chi_2}}{6} \ .\label{095152_5Mar20}
\end{equation}
We can thus see that if the first derivatives of the potential are bounded, the Hubble parameter and the anisotropy parameter are also bounded. By substituting \eqref{095152_5Mar20} into the Friedmann equation~\eqref{friedmann_K2}, we obtain a constraint equation for $\chi_1$ and $\chi_2$,
\begin{equation}
 \frac{1}{3}(1-3\chi_1)V_{\chi_1} - \frac{1}{2}(1+2\chi_2)V_{\chi_2}=V+\frac{\rho}{\mu^2 \Mpld} \ .\label{190042_26Apr20}
\end{equation}

We consider the vacuum case $\rho = p = p_{\pm} = 0$ from now on. The equations of motion for the anisotropies become
\begin{equation}
 \frac{\mathrm{d}}{\mathrm{d}t}\left[a^{3}(1 +2\chi_2 )\sigma_\pm\right]=0 \ ,
\end{equation}
which can be integrated to yield
\begin{equation}
 \sigma_\pm = \frac{\sigma_\pm^{(0)}}{(1+2\chi_2) a^{3}} \ .
\end{equation}
Consequently, we obtain
\begin{equation}
 \Sigma^2=\frac{\Sigma^2_0}{(1+2\chi_2)^2 a^6}\ .
\end{equation}
As an example, let us choose the following function as the limiting potential,
\begin{equation}
 V(\chi_1, \chi_2)=\chi_1-\tanh \chi_1 + \chi_2 -\tanh \chi_2\ .
 \label{limiting_potential_example}
\end{equation}
This function satisfies all the following requirements: (1) guaranteeing the finiteness of the Hubble parameter and finiteness of the anisotropy parameter at all energy scales; and (2) recovering Einstein gravity at low energies. We explain the explicit conditions put on the limiting potential in the \hyperref[ssec:limpot]{Appendix}. Explicitly, the first derivatives of the potential can be evaluated as
$V_{\chi_1} = \tanh^2 \chi_1$, $V_{\chi_2} = \tanh^2 \chi_{2}$, and therefore, they satisfy $0 \leq V_{\chi_1} < 1$, $0 \leq V_{\chi_2} < 1$. Thus, for this choice of potential, $H^2$ and $\Sigma^2$ are bounded as
\begin{equation}
 0\leq H^2 < \frac{\mu^2}{9}\ , \qquad 0\leq \Sigma^2 < \frac{\mu^2}{6} \ .
\end{equation}

\begin{figure*}
 \centering
 \includegraphics[scale=0.47]{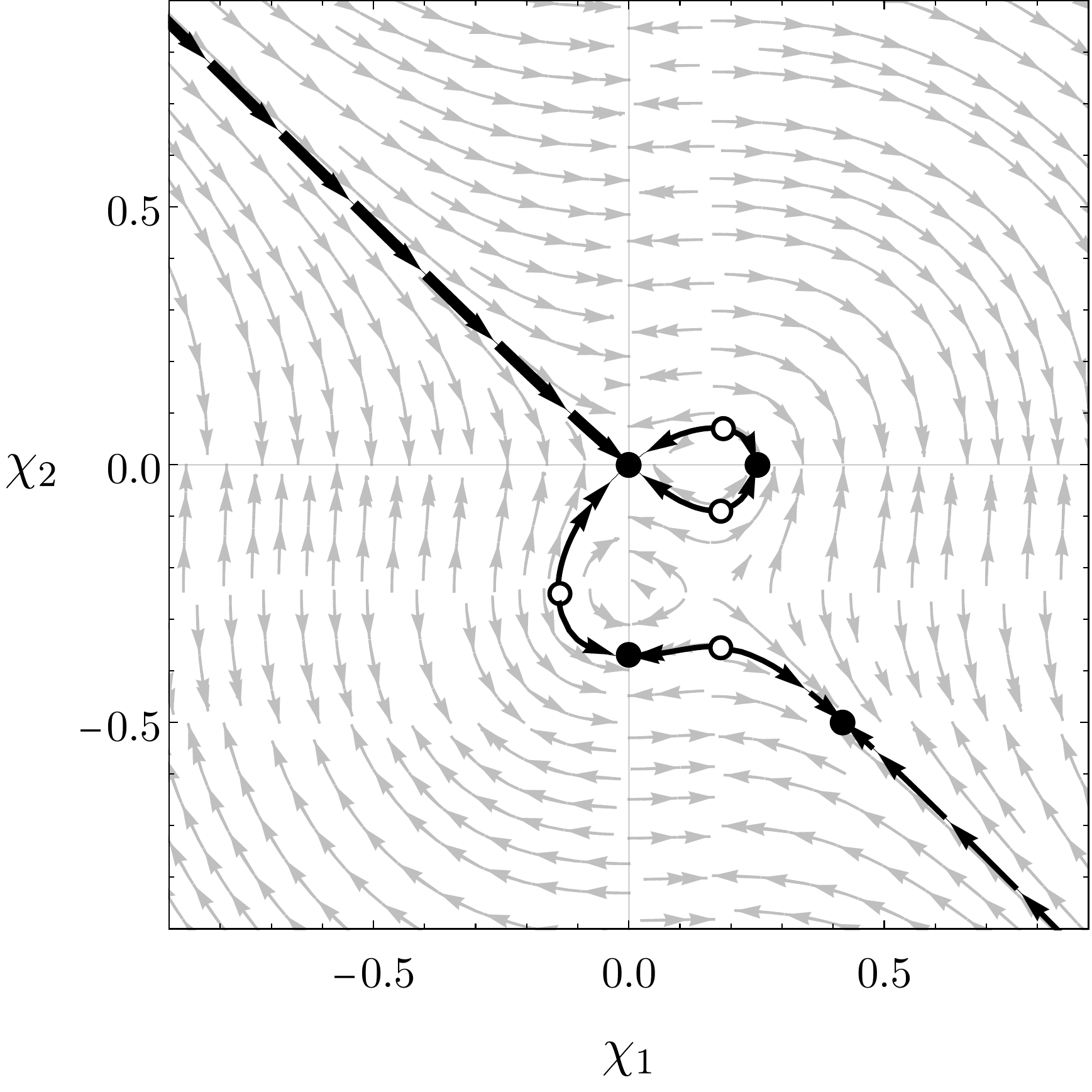}
 \hspace{5mm}
 \includegraphics[scale=0.47]{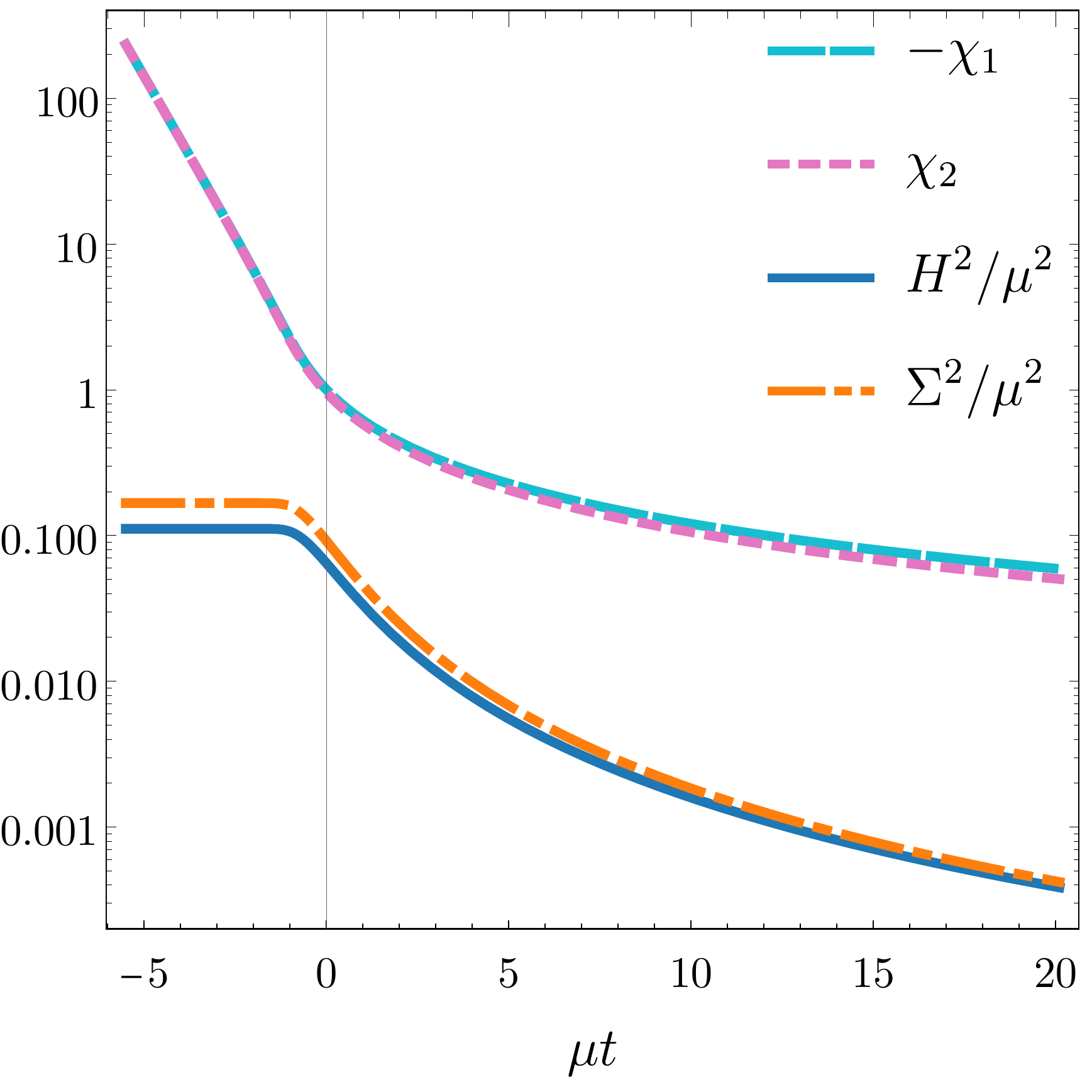}
 \caption{\textit{Left:} Phase diagram of the first-order differential equations~\eqref{chi1dot} and \eqref{chi2dot}. Only the black curves correspond to solutions of the constraint equation~\eqref{190042_26Apr20} with $\rho=0$. The filled and open circles represent stable and unstable fixed points, respectively. The thicker black curve in the upper left quadrant represents the solution that is numerically solved in the right panel.
 \textit{Right:} Numerical background solutions corresponding to the thick black curve in the left panel. The Hubble parameter~$H$ and the anisotropy parameter $\Sigma$ each reach their limiting value in the regime $\mu t < 0$.}
\label{fig:background}
\end{figure*}

In the left panel of Fig.~\ref{fig:background}, we show the possible evolution of the spacetime, constrained by the relation~\eqref{190042_26Apr20}, on the $(\chi_1,\chi_2)$-plane. Since $(\chi_1,\chi_2)=(0,0)$ is where the contribution of the limiting potential vanishes and Einstein gravity is recovered, the path starting from the upper left and terminating at the origin is the most desirable one. Provided that $H>0$, the time evolution is determined by
\begin{subequations}
\begin{align}
 \frac{\mathrm{d}\chi_1}{\mathrm{d}(\mu t)}&=  
 \frac{3(1+2\chi_2)V_{\chi_2}\sqrt{V_{\chi_1}}}{6V_{\chi_1}-(1-3\chi_1)V_{\chi_1 \chi_1}} 
 =(\sgn \chi_1) \frac{3(1+2\chi_2)\cosh^2{\chi_1}\tanh^2{\chi_2}}{-2+6\chi_1+3\sinh{2\chi_1}}\ ,\label{chi1dot}\\
  \frac{\mathrm{d}\chi_2}{\mathrm{d}(\mu t)}&=
  -\frac{2(1+2\chi_2)V_{\chi_2}\sqrt{V_{\chi_1}}}{4V_{\chi_2}+(1+2\chi_2)V_{\chi_2\chi_2}}
  =-\frac{(1+2\chi_2)\cosh{\chi_2}\sinh{\chi_2}|\tanh{\chi_1}|}{1+2\chi_2+\sinh{2\chi_2}}\ ,\label{chi2dot}
\end{align}
\end{subequations}
where we used $V_{\chi_1\chi_2}=0$ in the first equality of each equation. These equations are obtained from the time derivative of the limiting relations~\eqref{095152_5Mar20}. Along this path, we solved the dynamics as shown in the right panel of Fig.~\ref{fig:background}, where $t=0$ is characterized by $\chi_{1}(t = 0) = -1$. From the figure, we see that $H^2$ and $\Sigma^2$ are almost constant and asymptotically reach their upper bound values for $\mu t < 0$. Assuming $\beta_-\equiv 0$ for concreteness, we can approximate the evolution of the scale factor and the anisotropy for $\mu t \rightarrow -\infty$ as
\begin{equation}
 a(t) \simeq e^{\frac{1}{3} \mu t}\ , \qquad \beta_+(t) \simeq \pm \frac{1}{\sqrt{6}} \mu t \ .
\end{equation}
The sign of $\beta_+$ is determined by its initial condition. Thus, the very early stage of the universe in this model is effectively described by the metric,
\begin{equation}
 g_{\mu\nu}\mathrm{d}x^{\mu}\mathrm{d}x^{\nu} = - \mathrm{d}t^2 + e^{ 2 H_{x} t }\mathrm{d}x^2 + e^{ 2 H_{y} t}  \mathrm{d}y^2 + e^{2 H_{z} t}\mathrm{d}z^2 \ ,
\end{equation}
with
\begin{equation}
 H_{x} = H_{y} = \left( \frac{1}{3} \pm \frac{1}{\sqrt{6}} \right) \mu \ , \qquad
 H_{z} = \left( \frac{1}{3} \mp \frac{2}{\sqrt{6}} \right) \mu \ .       
\end{equation}
Since the cosmological time $t$ (i.e., the proper time for comoving observers) is defined all the way to $t \rightarrow - \infty$, the comoving time-like geodesics are past complete. Although null geodesics are expected to be past incomplete as in the case of the flat de Sitter universe (see Refs.~\cite{Borde:2001nh,Yoshida:2018ndv}), our formulation ensures that the past boundary is not a scalar curvature singularity at least up to $\mathcal{O}(\partial^2 g)$ because the curvature invariants~$R$, $R_{\mu\nu}R^{\mu\nu}$, and $R_{\mu\nu\rho\sigma}R^{\mu\nu\rho\sigma}$ approach constant values.

\subsection{Stability of the anisotropic background}

We examine the stability of the Bianchi I solution that we found in the previous subsection against perturbations. We note that because of the cuscuton-type construction of the theory, we have only two physical degrees of freedom corresponding to gravitational waves on an FLRW spacetime. For simplicity, we keep the rotational symmetry in the $xy$-plane for the background metric by setting $\beta_-\equiv 0$, namely, 
\begin{equation}
 g_{\mu\nu}\mathrm{d}x^{\mu}\mathrm{d}x^{\nu} = - \mathrm{d}t^2 + a(t)^{2} \left[ e^{2\beta(t) }(\mathrm{d}x^2 + \mathrm{d}y^2) + e^{-4 \beta(t)}\mathrm{d}z^2\right]  \ .
\end{equation}
From here on, we write $\beta\coloneqq \beta_+$ and $\sigma\coloneqq \dot{\beta}_+$. In this case, the perturbations can be categorized into vector perturbations and scalar perturbations, and they evolve independently at linear order. Thus, we investigate each type of perturbation separately.

\subsubsection{Vector perturbations}

Thanks to the rotational symmetry in the $xy$-plane, for a given Fourier mode of perturbation with wavevector $\mathbf{k}$ we can always choose the $x$- and $y$-axes so that $k_{i} \mathrm{d}x^{i} = k_{y} \mathrm{d}y + k_{z} \mathrm{d}z$. The easiest way to derive the second-order perturbed action for this mode is to assume that all the perturbation variables depend only on $(t,y,z)$ in position space. On this type of anisotropic background, there are three independent vector-type perturbations for the metric and one for the vector field. Because of the gauge degree of freedom, $\xi_\mu=(0,\xi_x, 0, 0)$, where $\xi_x=\xi_x(t,y,z)$, we can eliminate one of the variables, resulting in three independent variables. We use this gauge degree of freedom to express the vector-type perturbations as
\begin{equation}
 \delta g_{\mu\nu}=
 \begin{pmatrix}
 0 & \delta E & 0 & 0  \\
 \ast  & 0  & - a^2 e^{2\beta}\partial_z h_\times  & a^2 e^{-4\beta}\partial_y h_\times  \\
 0 & \ast  &0  &0 \\
 0 & \ast  &0  &0
 \end{pmatrix}
 \ , 
\end{equation}
where the symbols~$\ast$ represent symmetric components, and
\begin{equation}
 \delta A_\mu = (0, \delta A_x, 0, 0) \ .
\end{equation}
This choice of perturbation variables allows us to easily take the isotropic limit. Indeed, $h_\times$ corresponds to the cross-mode tensor perturbation in the isotropic case~\cite{Yoshida:2017swb}. By using the equations of motion for $\delta E$ and $\delta A_x$ to eliminate themselves, we obtain the second-order perturbed Lagrangian in Fourier space,
\begin{equation}
 \delta^2\mathcal{L}_{\rm V}=\frac{\Mpld}{2}k^2 a^3 e^{ -4\beta}(1+2\chi_2)\left[\dot{h}_{\times,-\ve{k}}\dot{h}_{\times,\ve{k}}-\left(\frac{k^2}{(1+2\chi_2) a^2}+36\frac{e^{2\beta}k_y^2 k_z^2}{k^4} \sigma^2 \right)h_{\times,-\ve{k}}h_{\times,\ve{k}}\right]\ ,\label{eq:LV2g}
\end{equation}
where $k^2(t) \coloneqq  e^{-2\beta}k_y^2 + e^{4\beta} k_z^2$. We can see that any ghost instability is avoided if
\begin{equation}
 1+2\chi_2 > 0 \ ,\label{184309_7Apr20} 
\end{equation}
and the same condition guarantees the absence of any gradient instability.

We note that, in the isotropic limit (i.e., for an FLRW background), the quadratic Lagrangian takes the form
\begin{equation}
 \left.\delta^2\mathcal{L}_{\rm V}\right|_{\text{FLRW}}=\frac{\Mpld}{2}k^2 a^3 (1+2\chi_2)\left[\dot{h}_{\times,-\ve{k}}\dot{h}_{\times,\ve{k}}-\frac{k^2}{(1+2\chi_2) a^2 }h_{\times,-\ve{k}}h_{\times,\ve{k}}\right]\ ,\label{212638_7Apr20}
\end{equation}
which is free of instabilities under the condition~\eqref{184309_7Apr20}. In the isotropic limit where $\Sigma^2\propto V_{\chi_2}$ vanishes, $\chi_2$ vanishes as well for our choice of potential~\eqref{limiting_potential_example}. Therefore, the above quadratic Lagrangian coincides with that of Einstein gravity as expected. This is also true for any limiting potential recovering Einstein gravity at low energies since we require $V_2 \sim \chi_2^{m_2}$ with $m_2>1$ around $\chi_2=0$ (see the \hyperref[ssec:limpot]{Appendix}). On the other hand, if we have some potential minima at $\chi_2 \neq 0$, the overall coefficient is different for the different minima.

\subsubsection{Scalar perturbations}

Again, to derive the action for the mode with $k_{i}\mathrm{d}x^{i} = k_{y} \mathrm{d}y + k_{z} \mathrm{d}z $, we assume all the perturbation variables depend only on $(t,y,z)$. We have seven independent scalar-type perturbations for the metric and three for the vector field. Moreover, one should take into account the perturbations of the three scalar fields, $\delta \lambda$, $\delta \chi_1$, and $\delta \chi_2$. Because of the gauge degrees of freedom, $\xi_\mu=(\xi_0, 0, \partial_y \xi, \xi_z)$, we can eliminate $3$ out of $13$ scalar perturbations. These gauge degrees of freedom enable us to express the scalar-type perturbations associated with the metric as
\begin{equation}
 \delta g_{\mu\nu}=
 \begin{pmatrix}
 -2\Phi & 0 & a (\partial_y B+e^{2\beta}\partial_z s) & a (\partial_z B - e^{-4\beta}\partial_y s) \\
 0 & - a^2(\partial_y^2+e^{6\beta}\partial_z^2) h_+ & 0 & 0 \\
 \ast & 0 & a^2 e^{6\beta}\partial_z^2 h_+ & - a^2\partial_y \partial_z h_+  \\
 \ast & 0 & \ast  & a^2 e^{-6\beta}\partial_y^2 h_+  
 \end{pmatrix}
\end{equation}
and those associated with the vector field as
\begin{equation}
 \delta A_\mu = (\delta A_0, 0, \partial_y \delta A, \delta A_z) \ .
\end{equation}
Note that $h_+$ amounts to the plus-mode tensor perturbation in the isotropic case~\cite{Yoshida:2017swb}. From the variation with respect to $\delta \lambda$, we obtain $\delta A_0 - \Phi=0$. After eliminating all appearances of $\delta A_0$ in the action with this equation and performing an integration by parts, we can eliminate all the derivatives on the variables other than $h_+$. With the definition~$\psi^a=(\Phi, B,s, \delta A,\delta A_z,\delta \chi_1, \delta \chi_2)$, we have
\begin{equation}
 \delta^2 \mathcal{L}_{\rm S}=\mathcal{A} \dot{h}_{+,\ve{k}}\dot{h}_{+,-\ve{k}}+\mathcal{C}_a \dot{h}_{+,\ve{k}}\psi^a_{-\ve{k}}+\mathcal{C}_a^\ast \dot{h}_{+,-\ve{k}}\psi^a_{\ve{k}}+\mathcal{E}_a h_{+,\ve{k}}\psi^a_{-\ve{k}}+\mathcal{E}_a^\ast h_{+,-\ve{k}}\psi^a_{\ve{k}}-\mathcal{M}h_{+,\ve{k}}h_{+,-\ve{k}}+\mathcal{J}_{ab}\psi^a_{\ve{k}}\psi^b_{-\ve{k}}\ ,
\end{equation}
where $\mathcal{A}$, $\mathcal{C}_a$, $\mathcal{E}_a$, $\mathcal{M}$, and $\mathcal{J}_{ab}$ are functions described by the background quantities, with $\mathcal{J}_{ab}=\mathcal{J}_{ba}^\ast$ and $\det \mathcal{J}_{ab}\neq 0$. We do not write down the explicit expressions for all these functions for now. Rather, we emphasize the methodology, and the final expressions will be shown below. After substituting the equations of motion for $\psi^a_{\ve{k}}$ and $\psi^a_{-\ve{k}}$, we obtain
\begin{equation}
 \delta^2 \mathcal{L}_{\rm S}=\mathsf{A}\dot{h}_{+,\ve{k}}\dot{h}_{+,-\ve{k}}-\bigl[\mathcal{M}+(\mathcal{J}^{-1})^{ab}\mathcal{E}_a\mathcal{E}^\ast_b-\dot{\mathsf{D}}\bigr] h_{+,\ve{k}}h_{+,-\ve{k}}\ ,
\end{equation}
where $\mathsf{A}=\mathcal{A}-(\mathcal{J}^{-1})^{ab}\mathcal{C}_{a}\mathcal{C}_b^\ast$ and  $\mathsf{D}=(\mathcal{J}^{-1})^{ab}\mathcal{C}_a\mathcal{E}^\ast_b$. Note that here we used the fact that $\mathsf{D}$ is real, found by evaluating the explicit form of $\mathsf{D}$.

At large $k$, the second-order perturbed Lagrangian takes the following form:
\begin{equation}
 \delta^2 \mathcal{L}_{\rm S}=\frac{\Mpld}{2} a^3 k^4 (1+2\chi_2)\bigl[\mathcal{G}\dot{h}_{+,\ve{k}}\dot{h}_{+,-\ve{k}}-\mathcal{K} h_{+,\ve{k}}h_{+,-\ve{k}}\bigr]\ .
\end{equation}
The $k$-dependent coefficients $\mathcal{G}$ and $\mathcal{K}$ are written as
\begin{equation}
 \mathcal{G}=\frac{\mathcal{G}_{\rm n}}{\mathcal{G}_{\rm d}}\ , \qquad
 \mathcal{K}=\frac{k^2}{(1+2\chi_2)a^2 }\ ,
 \label{coefficient_qLag}
\end{equation}
where
\begin{align}
 \mathcal{G}_{\rm n}&=6(1+\kappa^2)^2(1+2\chi_2)\chi_2^2 H \left[(3\chi_1+2\chi_2) \sigma^2- H\dot{\chi}_1\right] \nonumber \\
 &\quad +\left\{\left[3\kappa^2(3 \chi_1+2\chi_2)+(-2+\kappa^2)^2(-1+3\chi_1)\chi_2^2\right] \sigma^2-3\kappa^2 H \dot{\chi_1}\right\}\dot{\chi}_2\ , \\
 \mathcal{G}_{\rm d}&=\mathcal{G}_{\rm n}+3\kappa^4\left[(3\chi_1+2\chi_2) \sigma^2- H\dot{\chi}_1\right]\chi_2^2\dot{\chi}_2\ .
\end{align}
Here, we defined
\begin{equation}
 \kappa(t)\coloneqq \left|\frac{k_y}{k_z}\right| e^{-3\beta(t)}\ ,
\end{equation}
which is nothing but the ratio between the $y$- and $z$-components of the physical wavevector, i.e., $\kappa(t) = \hat{k}_{y}(t)/ \hat{k}_{z}(t)$. The physical wavenumbers are defined in terms of the components of $\ve{k}$ with respect to the tetrad basis:
\begin{align}
 k_{i}\mathrm{d}x^{i} = k_{y} \mathrm{d}y + k_{z} \mathrm{d}z \eqqcolon \hat{k}_{y}(t) \left( a(t) e^{\beta(t)} \mathrm{d}y \right) +  \hat{k}_{z}(t) \left( a(t) e^{-2 \beta(t)} \mathrm{d}z \right)\ .
\end{align}
We see that there is no gradient instability as long as the condition~\eqref{184309_7Apr20} is satisfied, though the condition for the absence of ghost instabilities is not obvious at this point.

On an FLRW background, the quadratic Lagrangian for the perturbations corresponding to the plus-mode gravitational waves is reduced to the following form,
\begin{equation}
 \left.\delta^2 \mathcal{L}_{\rm S} \right|_{\text{FLRW}}
 =\frac{\Mpld}{2} a^3k^4(1+2\chi_2)\biggl[\dot{h}_{+, \ve{k}}\dot{h}_{+,-\ve{k}}-\frac{k^2}{(1+2\chi_2) a^2}h_{+,\ve{k}}h_{+,-\ve{k}}\biggr]\ ,
\end{equation}
without taking the large-$k$ limit. This expression coincides with the one for the cross mode~\eqref{212638_7Apr20}. Thus, as long as the condition~\eqref{184309_7Apr20} is satisfied, the FLRW background is stable against both the plus- and cross-mode tensor perturbations up to the linear order. Once again, $\chi_2$ vanishes in the isotropic limit for our choice of potential \eqref{limiting_potential_example}, and the quadratic perturbed action reduces to that of Einstein gravity.

\begin{figure*}
    \centering
    \includegraphics[scale=0.3905]{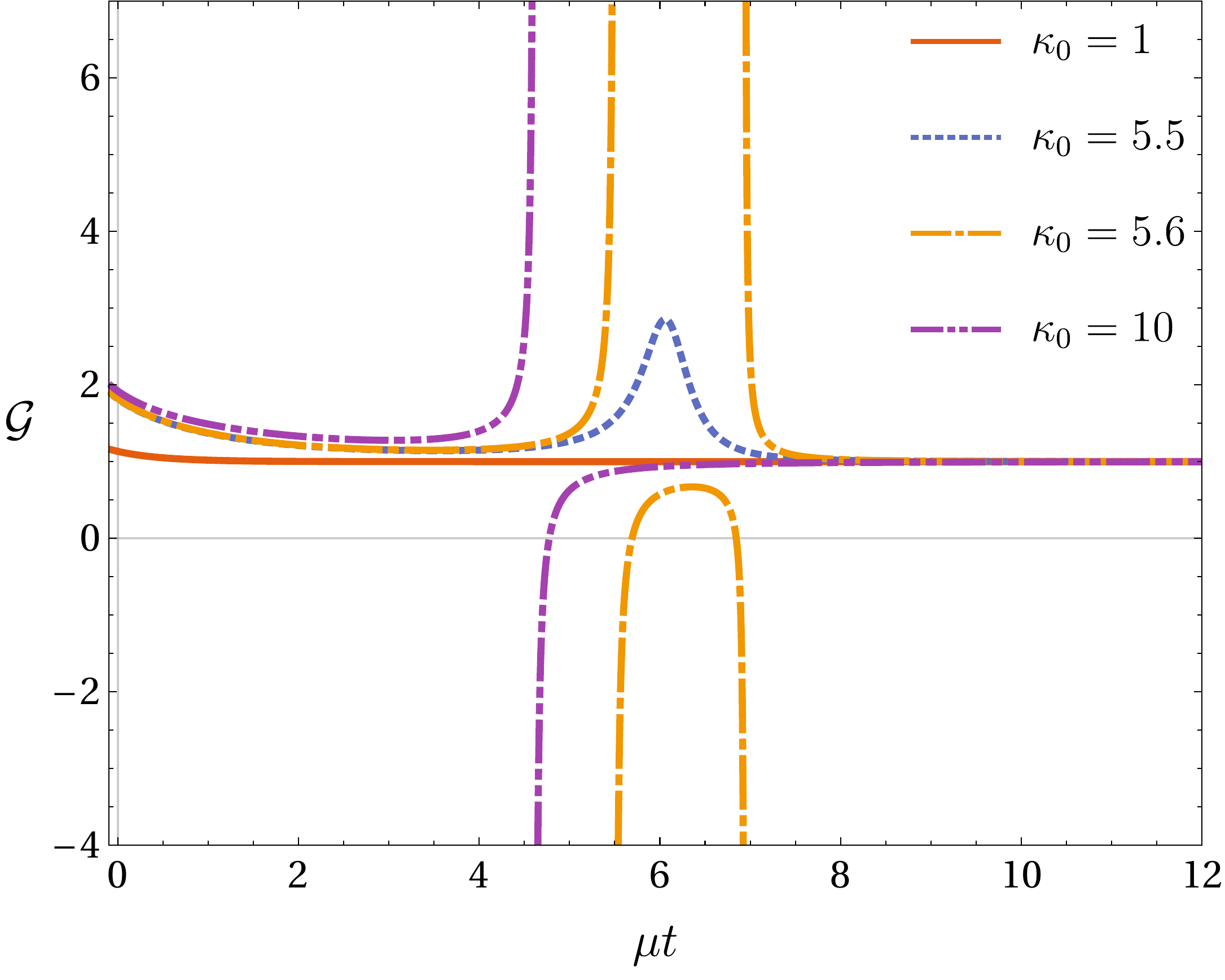}
    \hspace{5mm}
    \includegraphics[scale=0.3905]{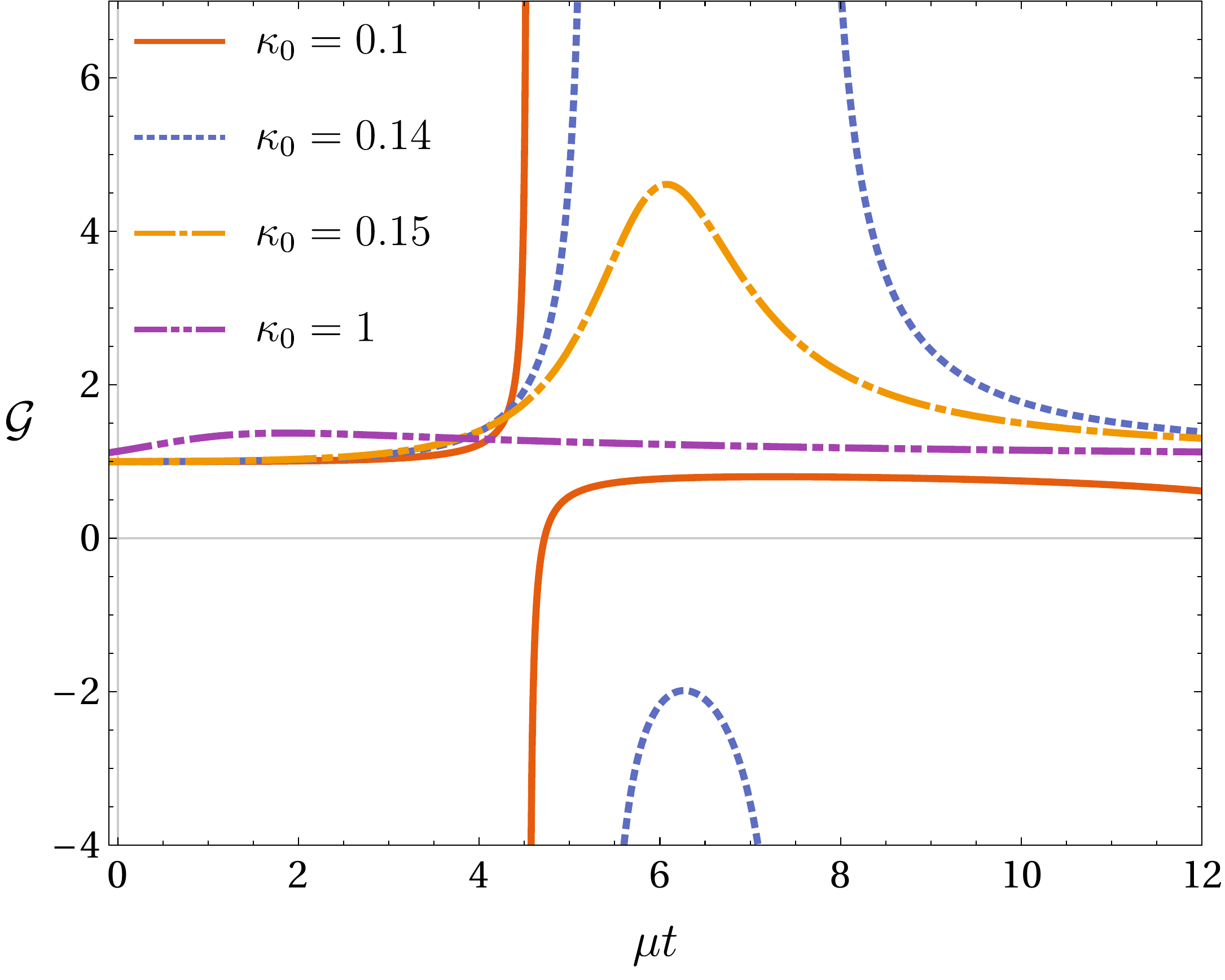}
    \caption{\textit{Left:} Time evolution of $\mathcal{G}$ for $\sigma > 0$. The behavior drastically changes near $\kappa_{0} \sim 5.5$. $\mathcal{G}$ is always positive for small-$\kappa_{0}$ modes with $\kappa_0 \lesssim 5.5$.
    \textit{Right:} Time evolution of $\mathcal{G}$ for $\sigma < 0$. The behavior drastically changes near $\kappa_{0} \sim 0.15$. $\mathcal{G}$ is always positive for large-$\kappa_{0}$ modes with $\kappa_0 \gtrsim 0.15$.}
    \label{fig:stability}
\end{figure*}

Let us go back to the anisotropic case and discuss under which situation ghost instabilities can be avoided. For simplicity, we focus on large-$k$ modes. The stability is guaranteed if $\mathcal{G}$ in \eqref{coefficient_qLag} is positive. The quantity~$\mathcal{G}$ can be expressed in terms of $\kappa$, $\chi_1$, and $\chi_2$ by making use of the background equations and the explicit potential form, i.e., $\mathcal{G}=\mathcal{G}(\kappa,\chi_1,\chi_2)$. Since the functional structure of $\mathcal{G}$ is quite involved and its sign may change time to time, we study the time evolution of $\mathcal{G}$ numerically. We choose the origin of time by the condition $\chi_{1}(t = 0) = -1$. In addition, we vary the value of $\kappa$ at $t=0$, $\kappa_{0} \coloneqq  \kappa(t = 0)$. The numerical results under this setup are shown in the left and right panels of Fig.~\ref{fig:stability} for $\sigma > 0$ and $\sigma < 0$, respectively (recall $\sigma=\dot\beta$ here).
Let us first discuss the case with $\sigma>0$. From the left panel of Fig.~\ref{fig:stability}, we can see that ghost instabilities are avoidable when
\begin{align}
 \kappa_0 \lesssim 5.5 \qquad (\text{for $\sigma > 0$})\ .
\end{align}
We emphasize that this condition is automatically satisfied if we assume the initial conditions for perturbations are provided sufficiently far in the past. This is because the value of $\kappa(t)$ is exponentially damped as time progresses:
\begin{align}
 \kappa(t) = e^{- 3 \beta(t)} \kappa_{0} \sim e^{ - \frac{3}{\sqrt{6}} \mu t} \kappa_{0}\ .
\end{align}
Hence, if we set every initial condition at a sufficiently early time~$t = t_{\text{i}} < 0$ (i.e., $\mu |t_\mathrm{i}| \gg 1$), the ghost-free condition for the wavenumber at $t = t_{\text{i}}$ can be understood as
\begin{align}
 \kappa(t_{\text{i}}) \lesssim e^{\frac{3}{\sqrt{6}} \mu |t_{\text{i}}|} \times 5.5 \rightarrow \infty \qquad (\mu t_{\text{i}} \rightarrow - \infty)\ ,
\end{align}
and thus 
a large class of wavenumbers
can satisfy the stability condition. A similar argument holds also for $\sigma < 0$. From the right panel of Fig.~\ref{fig:stability}, ghost instabilities are avoidable when
\begin{align}
 \kappa_{0} \gtrsim 0.15 \qquad (\text{for $\sigma < 0$})\ .
\end{align}
This condition is also naturally satisfied because now $\kappa(t)$ grows exponentially as time progresses. Thus, the stability condition is satisfied for 
a large class of wavenumbers
if we set their initial conditions at a sufficiently early time:
\begin{align}
 \kappa(t_{\text{i}}) > e^{- \frac{3}{\sqrt{6}} \mu |t_{\text{i}}|} \times 0.15 \rightarrow 0 \qquad (\mu t_{\text{i}} \rightarrow - \infty)\ .
\end{align}

\section{Summary and Discussion}
\label{sec:discussion}

In this work, we proposed the limiting extrinsic curvature theory as a new class of limiting curvature theories. The general actions of two specific models are given by \eqref{generalactions}. We showed that mimetic gravity and cuscuton gravity are both contained in this category, and they are actually equipped with a mechanism limiting the Hubble parameter on a homogeneous spacetime. However, limiting the Hubble parameter is not enough to obtain a non-singular universe when the spacetime is not isotropic. In the context of the framework developed in this work, we constructed a minimal model limiting anisotropies by introducing an additional limiting potential for the anisotropies. For this model, we found a non-singular Bianchi I solution in the sense that there is no scalar curvature singularity. It starts from a phase of constant Hubble parameter and constant anisotropy parameter, and in vacuum, it ends up with Minkowski spacetime in the asymptotic future. We derived the stability conditions under the $SO(2)$ symmetry of the spacetime. Note that, as in cuscuton gravity, the theory in vacuum has only two dynamical degrees of freedom, $h_\times$ and $h_+$, i.e., counterparts of the cross- and plus-mode tensor perturbations on an isotropic spacetime, respectively. As far as the condition~$1+2\chi_2>0$ is satisfied, where $\chi_2$ is the auxiliary scalar field ensuring the boundedness of anisotropies, both modes are free of gradient instabilities. Moreover, ghost instabilities are absent for $h_{\times}$. For the $h_+$ mode, one can circumvent the ghost instabilities for 
a large class of wavenumbers
if the limiting phase lasts long enough, i.e., if we put the initial conditions for the perturbations much before the end of the limiting phase.
While this is a reasonable assumption, ghost instabilities may remain for arbitrary initial conditions.
In other words, there remains some small region of phase space where the model is unstable, and as a whole, one cannot claim full stability.

Though we analyzed the vacuum case where the spacetime approaches Minkowski space, it is straightforward to introduce a cosmological constant and/or matter fields to the theory. In this case, our framework can be understood as the early-time completion of the inflationary scenario without causing any inconsistency with experimental results in the low-energy regime. Yet, it would be interesting to see how adding matter might affect the stability of the cosmological perturbations. Another caveat is that we studied the case where the limiting potential has the form~$V(\chi_1,\chi_2)=V_1(\chi_1)+V_2(\chi_2)$ and $V_1(\chi)=V_2(\chi)$. If we introduce a hierarchy between $V_1$ and $V_2$, we can keep the anisotropy parameter smaller than the Hubble parameter at all times. However, in that case, we expect instabilities to appear in a much broader region of wavenumbers for the $h_+$ mode, which cannot be overcome by setting the initial conditions early.

Another limitation of the current model is that the auxiliary fields still grow without bound in the asymptotic past. Consequently, one must interpret the theory as an effective field theory whose regime of validity cannot fully include the limit $\chi_k\rightarrow\infty$. Determining the strong coupling scale would thus be an interesting follow-up (in the spirit of, e.g., Refs.~\cite{Koehn:2015vvy,deRham:2017aoj}). This may also have implications for how and when one may set the initial conditions for perturbations in the asymptotic past. This is certainly an issue that deserves a closer investigation, especially in the context of an anisotropic universe since an anisotropic spacetime is not conformally flat and the initial state would be different from the standard `Minkowski limit' Bunch-Davies state.

It is interesting to mention that our formalism can be extended to include the acceleration~$a_\mu\coloneqq n^\nu \nabla_\nu n_\mu$ of the spatial hypersurface. That is,
\begin{equation}
 S=\int \mathrm{d}^4 x\, \sqrt{-g}\left[\frac{\Mpld}{2}R+\lambda (A_\mu A^\mu+1)+\chi_1 (\nabla^\mu A_\mu)^2+\chi_2 \nabla^\mu A_\nu \nabla^\nu A_\mu + \chi_3 A^\nu \nabla_\nu A_\mu A^\lambda \nabla_\lambda A^\mu - V(\chi_1,\chi_2, \chi_3)\right]\ .
\end{equation}
We can interpret this action as a non-linear extension of Einstein-aether theory.\footnote{A subclass of Einstein-aether theory, precisely corresponding to the original cuscuton theory, was considered in Ref.~\cite{Casalino:2020vhl} to study a non-singular bouncing background, reproducing the dynamics of Loop Quantum Cosmology.}\ Indeed, by integrating out the auxiliary fields, we obtain
\begin{equation}
 S=\int \mathrm{d}^4 x\, \sqrt{-g}\left[\frac{\Mpld}{2}R+\lambda (A_\mu A^\mu+1)-f((\nabla^\mu A_\mu)^2,\nabla^\mu A_\nu \nabla^\nu A_\mu, A^\nu \nabla_\nu A_\mu A^\lambda \nabla_\lambda A^\mu)\right]\ ,
\end{equation}
where $f$ is a scalar function determined by the potential~$V$. A more direct relation can be seen with a potential of the form~$V=(\mu^2/2\Mpld) \sum_{k=1}^{3} (\chi_k-c_k \Mpld )^2$, where $\mu$ is a mass scale and the $c_k$'s are dimensionless parameters, though this is not a limiting potential. Expanded about $E/\mu \ll 1$, where $E^2\sim\max\{(\nabla^\mu A_\mu)^2,\nabla^\mu A_\nu \nabla^\nu A_\mu, A^\nu \nabla_\nu A_\mu A^\lambda \nabla_\lambda A^\mu\}$, this action is actually reduced to Einstein-aether theory with higher-order corrections~(see Eq.~(1) in Ref.~\cite{Zlosnik:2006zu} for a comparison). On the other hand, by identifying $A_\mu$ as the unit normal vector~$n_\mu$, the extrinsic curvature is written as $K_{\mu\nu}=\nabla_\mu A_\nu+A_\mu a_\nu$, which allows us to rewrite the above action as
\begin{equation}
 S=\int \mathrm{d}^4 x\, \sqrt{-g}\left[\frac{\Mpld}{2}R+\lambda (A_\mu A^\mu+1)+\chi_1 K^2+\chi_2 K^\mu{}_\nu K^\nu{}_\mu + \chi_3 a^\mu a_\mu - V(\chi_1,\chi_2, \chi_3)\right]\ .
\end{equation}
This gives us the picture that the non-linear extension of Einstein-aether theory includes a theory limiting the extrinsic curvature and the acceleration. Note that the number of dynamical degrees of freedom is five in general in the Einstein-aether theory, three of which disappearing when restricting ourselves to theories without the acceleration, corresponding to the kinetic term of the vector field~\cite{Jacobson:2004ts}.

Another interesting link can be made with Ref.~\cite{Ito:2019ztb}, where a Kaluza-Klein scenario was proposed within cuscuton gravity. In this scenario, a higher-dimensional spacetime can dynamically reduce to a four-dimensional inflationary spacetime with stable extra dimensions. It can thus be understood as a kind of anisotropic inflation in higher-dimensional spacetime. As such, the limiting anisotropy mechanism that was introduced in the present paper may be applied to obtain non-singular spacetimes in higher-dimensional theories. The theory presented in the present paper could also potentially be used to construct general anisotropic inflationary models by having the limiting anisotropy scale comparable to the inflationary energy scale. In such a context, the anisotropies present during inflation could leave specific imprints in the observable cosmological perturbations, and it would be interesting to see how these signals differ from those of `standard' anisotropic inflation models (see, e.g., Refs.~\cite{Gumrukcuoglu:2007bx,Pitrou:2008gk,Watanabe:2009ct,Dulaney:2010sq,Gumrukcuoglu:2010yc,Watanabe:2010fh,Soda:2012zm}).

Another context in which the present work may be interesting to apply is with regard to the initial conditions of the Universe. It was found in Ref.~\cite{Lehners:2019ibe} that spacetimes dominated by anisotropies in the approach to the big bang in the very early Universe tend to have a divergent action, indicating ill-defined path integrals and quantum amplitudes in the context of quantum cosmology. Accordingly, it was found that essentially only isotropic and accelerating spacetimes could originate from the big bang. In the present work, the `big bang' (the moment the spatial hypersurface reaches zero volume) is pushed to $t=-\infty$, and the presence of anisotropies would still allow for a convergent action since they are bounded (as is the Hubble parameter). Thus, within the model developed in the present paper, a constant-anisotropy and constant-Hubble parameter initial phase for the universe could be allowed under the principle of a finite action in the past. However, if the spacetime is extendible beyond the point where $a\rightarrow 0$ (as $t\rightarrow -\infty$) as explored in Ref.~\cite{Yoshida:2018ndv} for homogeneous and isotropic (quasi-)de Sitter spacetimes, then the full spacetime might have a previous contracting phase or have a cyclic past extension, in which case the past action could potentially diverge again. The conditions for extendibility of a spacetime with past null boundary are not known though when the assumption of isotropy is dropped.

Finally, an immediate follow-up to this work pertains to non-singular bouncing cosmology. As already mentioned, a homogeneous and isotropic bounce is straightforward to achieve within mimetic gravity or cuscuton gravity, and in the latter case, linear inhomogeneities have been shown to present no instability. Thus, the inclusion of anisotropies in the context of the cuscuton-type models developed in the present paper and studying their evolution through a bounce (in a similar fashion to Ref.~\cite{deCesare:2019suk}, which did the analysis for mimetic gravity) would be very interesting. If anisotropies are bounded in the same way as the Hubble parameter is, then it would imply that the BKL instability in the contracting phase (the rapid, chaotic blow up of the anisotropies) is evaded. There would remain to also study the evolution of perturbations to check whether or not the linear stability about an isotropic background, shown to hold in a cuscuton bounce, is spoiled when introducing anisotropies in addition to inhomogeneities.

\begin{acknowledgments}
Y.S.\ is supported by Young Teachers Training Program of Sun Yat-Sen University with Grant No.~20lgpy168 and thanks Kobe University for the long-term hospitality during the recent epidemic period.
D.Y.\ is supported by the JSPS Postdoctoral Fellowships No.\ 201900294 and the JSPS KAKENHI Grant Numbers 19J00294 and 20K14469.
Y.S.\ and D.Y.\ thank Jean-Luc Lehners and the Albert Einstein Institute (AEI) for hospitality during the beginning of this work.
Research at the AEI is supported by the European Research Council (ERC) in the form of the ERC Consolidator Grant CoG 772295 ``Qosmology''.
J.Q.\ further acknowledges financial support in part from the \textit{Fond de recherche du Qu\'ebec --- Nature et technologies} postdoctoral research scholarship and the Natural Sciences and Engineering Research Council of Canada Postdoctoral Fellowship.
J.Q.\ also thanks Jean-Luc Lehners for insightful discussions.
\end{acknowledgments}

\appendix*

\section{Appropriate choice of limiting potential for recovering Einstein gravity at low energies}
\label{ssec:limpot}

Here, we mention how to choose the potential function in limiting extrinsic curvature theories of the form
\begin{equation}
 S^{\rm (lim)}=\Mpld\int\mathrm{d}^4 x\, \sqrt{-g} \left[\sum_{k=1}^{n} \chi_k I_{k}(K_{\mu\nu},h_{\mu\nu},D_\mu)-\mu^2 V(\{\chi_k\})\right]\ ,
\end{equation}
such that the total action is the sum of the Einstein-Hilbert action, the matter action, and the above limiting curvature action. In the above, the $I_k$'s are functions having mass dimension two and $\mu$ is a mass parameter characterizing the potential. For simplicity, we assume the potential term can be separated into $n$ functions as $V(\{\chi_k\}) \coloneqq V(\chi_1,\dots,\chi_n) = \sum_{k=1}^{n} V_k(\chi_k)$. We also assume that the $\chi_k$'s have large absolute values at high energies and small ones at low energies, namely, the $\chi_k$'s are expressed in terms of positive powers of the energy scale asymptotically. To limit the extrinsic curvature, we require the potential at high energies to behave at most linearly, i.e., for each $k$, we want
\begin{equation}
 V_{k}(\chi_{k})\sim \mathcal{O} (\chi_{k})  \qquad \text{as} \; \chi_k\rightarrow\infty \ .
\end{equation}
Of course, we require that the first derivatives of the potentials, $\partial V_k/\partial\chi_k$, are finite for any field values of $\chi_{k}$ as well. On the other hand, as far as one thinks of the limiting curvature mechanism as coming from quantum corrections at high energies, we need to recover Einstein gravity when the curvature is small. This means the corrections should have a higher mass dimension than that of Einstein gravity. If the potentials behave as power laws for small $\chi_{k}$, i.e.,
\begin{equation}
 V_{k}(\chi_{k})\sim \chi_{k}^{m_{k}} \qquad \text{as}\; \chi_k\rightarrow 0\ ,
\end{equation}
where the $m_k$'s are real numbers, the curvature invariants scale as $I_k=\partial V_k/\partial\chi_k\sim \mu^2 \chi_k^{m_k-1}$. The correction terms in the action then scale as
\begin{equation}
 \chi_k I_k \sim \mu^2 V_k \sim \mu^2\left(\frac{I_k}{\mu^2}\right)^{m_k/(m_k-1)}\ .
\end{equation}
Since $I_k$ has mass dimension two, $R$ and $I_k$ should be of the same order. Correspondingly, the ratio of the quantum corrections to the Ricci scalar is evaluated as
\begin{equation}
 \frac{\chi_k I_k}{R}\sim \frac{\mu^2 V_k}{R}\sim \left(\frac{I_k}{\mu^2}\right)^{1/(m_k-1)} \sim \left(\frac{\rho_{\rm matter }}{\mu^2 \Mpld}\right)^{1/(m_k-1)}\ ,
\end{equation}
where we used $I_k\sim R\sim \rho_{\rm matter}/\Mpld$. Therefore, if we require
\begin{equation}
 m_k> 1
\end{equation}
for each $k$, we can recover Einstein gravity at low energies with $\rho_{\rm matter }/(\mu^2 \Mpld)\ll 1$.

In summary, provided that the potential function~$V$ is separable into $n$ functions of $\chi_k$, one should fix the potential such that $V_k(\chi_k)\sim \mathcal{O}(\chi_k)$ as $\chi_k\to \infty$ and $V_k(\chi_k)\sim \chi_k^{m_k}$ as $\chi_k\to 0$, with $m_k>1$. As a concrete example, we chose a potential function satisfying these two requirements in \eqref{limiting_potential_example}. Indeed, for $V_k(\chi_k)\propto \chi_k-\tanh\chi_k$, we have $V_k(\chi_k)\sim \chi_k$ as $\chi_k\to \infty$ and $V_k(\chi_k)\sim \chi_k^3$ as $\chi_k\to 0$.

\bibliographystyle{mybibstyle}
%

\end{document}